\documentclass[conference]{IEEEtran}
\IEEEoverridecommandlockouts
\usepackage{cite}
\usepackage{amsmath,amssymb,amsfonts}
\usepackage{algorithmic}
\usepackage{graphicx}
\usepackage{textcomp}
\usepackage{xcolor}
\usepackage[table]{xcolor}

\newcommand{\celllb}[1]{\cellcolor{blue!15}\textbf{#1}} 
\newcommand{\celldb}[1]{\cellcolor{blue!35}\textbf{#1}} 

\usepackage{url}
\usepackage{booktabs}
\def\BibTeX{{\rm B\kern-.05em{\sc i\kern-.025em b}\kern-.08em
    T\kern-.1667em\lower.7ex\hbox{E}\kern-.125emX}}
\begin{document}

\title{\textbf{\texttt{RoboKA:}} KAN Informed Multimodal Learning for RoboCall Surveillance System}

\author{
\IEEEauthorblockN{
Nitin Choudhury\IEEEauthorrefmark{1}, 
Nikhil Kumar\IEEEauthorrefmark{1}, 
Aditya Kumar Sinha\IEEEauthorrefmark{1}, 
Abhijeet Anand\IEEEauthorrefmark{1},
Hossein Salemi\IEEEauthorrefmark{2}, \\
Orchid Chetia Phukan\IEEEauthorrefmark{1}, 
Hemant Purohit\IEEEauthorrefmark{2}, 
Arun Balaji Buduru\IEEEauthorrefmark{1}
}
\IEEEauthorblockA{
\IEEEauthorrefmark{1}Indraprastha Institute of Information Technology Delhi, New Delhi, India\\
\IEEEauthorrefmark{2}George Mason University, Fairfax, VA, USA\\
Corresponding author: nitinc@iiitd.ac.in
}
}

\maketitle 

\begin{abstract}

Wide exploration on robocall surveillance research is hindered due to limited access to public datasets, due to privacy concerns. In this work, we first curate \textbf{\texttt{Robo-SAr}}, a synthetic robocall dataset designed for robocall surveillance research. \textbf{\texttt{Robo-SAr}} comprises of $\sim$1200 unwanted and $\sim$1200 legitimate synthetic robocall samples across three realistic adversarial axes: psycholinguistics-manipulated transcripts, emotion-eliciting speech, and cloned voices. We further propose \textbf{\texttt{RoboKA}}, a Kolmogorov–Arnold Network (KAN)-based multimodal fusion framework designed to model structured nonlinear interactions between acoustic and linguistic cues that characterize diverse adversarial robocall strategies. \textbf{\texttt{RoboKA}} first leverages cross-modal contrastive learning to align latent modality representations and feeds the resulting embeddings to a KAN-projection head for final classification. We benchmark \textbf{\texttt{RoboKA}} against strong unimodal and multimodal baselines in both in-domain and out-of-domain setups, finding \textbf{\texttt{RoboKA}} to surpass all baselines in terms of recall and F1-score.

 
\end{abstract}


\section{Introduction and Related Work}

Robocalls are defined as the automated calls generated using text-to-speech (TTS) systems or pre-recorded audio, primarily designed to share information with a large audience in an efficient manner~\cite{fcc-2, dmc-robocall}. These calls are being utilized across various sectors, including healthcare, customer service, and local government, due to their reliability, scalability, and cost efficiency. However, the widespread availability of such systems leads malicious users to craft deceptive calls, misleading users into revealing sensitive information or making fraudulent payments. Recent advancements in neural TTS models have further intensified this threat, increasing the quality, expressiveness, and naturalness of such deceptive robocalls. In parallel, large language models (LLMs) have demonstrated remarkable capabilities in generating linguistically sophisticated and psychologically persuasive content. In particular, with access to these advanced TTS and LLM models, adversaries can now leverage such systems, combined with LLMs, to generate highly deceptive robocalls that are difficult to distinguish from legitimate ones~\cite {vishing}. In recent years, i.e., 2024 and 2025, an approximate volume of 96.8 billion robocalls was reported in the USA, resulting in a massive loss of approximately 28.6 billion USD~\cite{prnews-robocall, yahoo-robocall}, demonstrating essential need for effective robocall surveillance systems.

Addressing the growing threat of robocalls, the regulatory framework STIR/SHAKEN has been introduced, and it has shown effectiveness against certain attack vectors~\cite{ftc-stir-shaken, stir-shaken}. However, they authenticate caller identities to reduce call spoofing; they fall short in addressing robocalls originating from international sources or exploiting infrastructure loopholes ~\cite{ftc-stir-shaken, prasad2024characterizingrobocallsmultiplevantage}. Similarly, network-based and behavior-based surveillance systems frequently fail against evasive and dynamic tactics used by advanced robocall campaigns. In response to these challenges, researchers have explored audio and text modalities, leveraging conventional acoustic features as well as pretrained language and speech model representations for the task of identifying such deceptive robocalls. Experimental results indicate that unimodal detection systems—based solely on audio or transcript—can achieve strong performance in controlled settings~\cite{elizalde2021detection, prasad2024characterizingrobocallsmultiplevantage, pandit2023combating}. \textit{Despite exploration and claims, most of the research suffers from critical limitations: their datasets are either proprietary, small-scale, or not publicly released, making reproducibility difficult and raising concerns about the reliability of reported results. In contrast, the robustness of such defensive systems under distributional shifts remains unexplored. This lack of reproducible benchmarks hinders cumulative progress in robocall defense research.} Unimodal robocall detectors are inherently fragile, as attackers can independently manipulate either speech acoustics or linguistic content, whereas jointly reasoning over audio and text constrains such evasions and is therefore necessary for robust detection.

Addressing the issue of reproducibility, this work curates and releases \textbf{\texttt{Robo-SAr}} (\textbf{\underline{Robo}Call--\underline{S}imulated \underline{A}dversa\underline{r}ial dataset)}, a novel adversarially enriched synthetic corpus. \textbf{\texttt{Robo-SAr}} targets three adversarial factors in transcript and speech synthesis: (i) LLM-driven psycholinguistic manipulation, (ii) TTS-driven emotion elicitation, and (iii) TTS-driven voice cloning. We use a large language model to curate and manipulate the psycholinguistics of the transcripts, and four widely used open-access neural text-to-speech (TTS) models with fourteen distinct voices to synthesize speech. Among these, we elicit eight emotions with two voices. In addition, we incorporate a real-world robocall dataset from the FTC Do Not Call repository~\cite{robocallDataset}, which contains real-world, genuine unwanted calls. Together, these components capture both plausible adversarial robocalls and real-world distribution shifts, enabling a robust and reproducible benchmark. Building on this, we propose \textbf{\texttt{RoboKA}} (\textbf{Unwanted \underline{Robo}Call Detection guided by \underline{KA}N}), a multimodal framework that (i) uses Cross-Modal Contrastive Learning (CMCL) on modality-specific pretrained embeddings to align audio and text representations and make them more label-consistent and robust under shifts, and (ii) replaces linear/MLP heads with modality-specific KAN projection heads plus stacked KAN fusion layers to capture nonlinear interactions and sharpen the boundary between unwanted and legitimate robocalls. \textit{We hypothesize that robocall adversaries implies modality-specific noise, requiring fusion models to perform conditional calibration and expressive nonlinear interaction modeling, which MLP fusion often lacks to preserve. Addressing this, we propose a KAN-based fusion approach to enable more stable evidence aggregation via expressive non-linear interactions.}



To summarize, the contributions of this work are as follows:

\begin{itemize}
    \item We curate and release \textbf{\texttt{Robo-SAr}}, a novel, adversarially enriched synthetic robocall corpus that explicitly targets three attack axes in both transcript and speech generation: psycholinguistic manipulation, emotion elicitation, and voice cloning, using LLM-based transcript design and four neural TTS models with fourteen voices and eight emotions, and is validated by expert annotators\footnote{We commit to releasing code and data upon reviews in compliance with ethical standards.}.


    \item We propose \textbf{\texttt{RoboKA}}, a multimodal framework that applies cross-modal contrastive learning to modality-specific pre-trained downstream embeddings and refines them via two KAN-based projection heads to enhance nonlinear expressiveness. 
    To our knowledge, we are the first to leverage KAN nonlinear projection and fusion for multimodal robocall detection under adversarial distribution shifts.

    \item We demonstrate that \textbf{\texttt{RoboKA}} yields more discriminative decision boundaries between unwanted and legitimate robocalls and consistently outperforms unimodal and multimodal baselines under both in-domain (InD) and out-of-domain (OoD) evaluations.
\end{itemize}

\section{Background}

\subsection{Text-to-Speech (TTS)}

TTS systems are a core component of speech synthesis, converting written text into human-audible speech~\cite{nvidia-tts}. Based on previous literature~\cite{wang2024evaluating-bark, hu2024can-openai, deng2025indextts-xtts, deng2023prosody-speecht5}, four state-of-the-art TTS systems: \texttt{Bark}~\cite{bark}, \texttt{OpenAI TTS}~\cite{openai2024tts}, \texttt{SpeechT5}~\cite{speecht5}, and \texttt{xTTS}~\cite{coqui2023xtts} are chosen for speech synthesis purpose. A total of fourteen voice samples from both male and female are considered for the speech synthesis process.

\subsection{Large Language Model (LLM)}

Recent advances in large language models have demonstrated higher linguistic manipulation capabilities. In particular, adversaries can now leverage these LLMs, combined with advanced TTS, to generate highly deceptive robocalls. In this work, we consider ChatGPT by OpenAI~\cite{chatgpt} for psycholinguistic manipulation as it demonstrates superior performance in linguistic manipulation in previous literature~\cite{sison2024chatgpt-chatgpt, zhan2023deceptive-chatgpt, figueiredo2024feasibility-chatgpt}.

\subsection{Pre-trained Models (PTMs)}

\textbf{\textit{Audio PTMs.}} 
We utilize three audio PTMs: Wav2Vec2~\cite{wav2vec2}, WavLM~\cite{wavlm}, and HuBERT~\cite{hubert} for modeling raw audio waveforms. Wav2Vec2 is trained in a self-supervised manner on raw audio using a contrastive objective, enabling it to learn stable speech features without requiring labels. WavLM follows a similar setup but adds masked prediction and denoising to handle varied and noisy speech. HuBERT employs a BERT-style masked prediction method applied to clustered acoustic units, enabling it to capture both phonetic and rhythmic structure. For all three, we use the base versions from Huggingface and extract 768-dimensional feature vectors from the frozen models’ last hidden layers using mean pooling.

\textbf{\textit{Text PTMs.}} 
In parallel, we utilize three text PTMs, including BERT~\cite{bert}, RoBERTa~\cite{liu2019roberta}, and GPT-2~\cite{gpt2}. BERT is trained with a masked language modeling objective (and, in the original setup, next-sentence prediction) on large text corpora, forcing it to infer missing tokens from context and yielding strong bidirectional sentence representations. RoBERTa retains the same core Transformer encoder architecture and masked language modeling objective, but eliminates next-sentence prediction and relies on heavier pretraining (utilizing more data, longer training, and dynamic masking), which typically yields more robust and transferable text features. GPT-2 utilizes a Transformer decoder trained with a causal (next-token) language modeling objective, learning to predict each token based solely on the previous context. This makes it naturally suited for generative modeling and autoregressive text representations. For all three, we use the base checkpoints from Hugging Face and extract 768-dimensional feature vectors from the frozen models’ last hidden layers using mean pooling.

\textbf{\textit{Kolmogorov-Arnold Network (KAN).}} A Kolmogorov-Arnold Network (KAN) is a neural architecture that replaces the usual ``linear layer + activation'' pattern with learnable functions on the connections~\cite{liu2024kan}. Instead of computing each neuron as a weighted sum followed by a nonlinearity, a KAN models the output as a sum of univariate functions applied to individual inputs (often implemented with spline-parameterized edge functions). This makes KANs expressive while staying relatively parameter-efficient, and it can improve interpretability because each input-to-output contribution can be inspected through its learned univariate function. In practice, KANs are used as drop-in alternatives to MLP blocks for function approximation and feature transformation, sometimes showing better data efficiency or robustness when the target relationship is well captured by structured, low-dimensional nonlinearities.

\section{Dataset: \textbf{\texttt{Robo-SAr}}}
\label{sec:dataset}

\textit{\textbf{Transcript Curation.}} We first curate 1233 unwanted and 1230 legitimate distinct call transcripts to construct \textbf{\texttt{Robo-SAr}}. We remove personally identifiable information (PII) and then leverage \texttt{ChatGPT-4o}~\cite{chatgpt} to curate and augment the transcripts, generating robocall-like content and injecting psycholinguistic manipulations, such as urgency, empathy, and authority framing, without modifying the hidden intent. The corpus is seeded on the Fraud Call India corpus\footnote{\url{https://www.kaggle.com/datasets/narayanyadav/fraud-call-india-dataset}} following previous literature~\cite{nicholas2024scamdetector, han2025scamgen}.

\textit{\textbf{Speech Synthesis.}} We leverage four advanced TTS engines, namely, \texttt{Bark}, \texttt{OpenAI TTS}, \texttt{SpeechT5}, and \texttt{xTTS} to synthesize two adversarially enriched speech modes: (i) \emph{voice cloning}, using four well-known Indian celebrity voices (two male, two female) and eight OpenAI TTS voices (five male, three female) to simulate impersonation attacks and increase speaker diversity; and (ii) \emph{emotion elicitation}, where emotion labels predicted by \texttt{roberta-base-conv-emotion}~\cite{roberta-base-conv-emotion} (eight categories: \textit{Surprised, Angry, Sad, Joyful, Anxious, Hopeful, Confident}, and \textit{Disappointed}) were passed to xTTS, enabling dynamic emotion-controlled prosody to strengthen adversarial realism. We leverage the default "en-us" for the speaking accent.

\textit{\textbf{Call Simulation.}} All the synthesized samples are then band-passed in the range of 300-3400 Hz, reflecting \textit{standard telephony voiceband and augmented, and also cross-validating with the RoboCall corpus from the Do Not Call Registry by FTC}~\cite{robocallDataset}, with light white noise (24-26 dB SNR) to simulate line artifacts~\cite{cisco, bandpass}.

\textit{\textbf{Data Validation.}} We evaluate the quality of the dataset in two ways: (i) transcript reconstruction correctness via word error rate (WER) calculation, and (ii) human validation. 

(i) We leverage OpenAI Whisper~\cite{radford2022whisper} for transcribing the synthesized speech \textit{before the call simulation} phase. For all the samples, we calculate the WER and found the maximum error rate to be $\sim$7\% and an average WER of $\sim$4\%.

(ii) We evaluate realism and quality \textit{after call simulation} via a human study with five experienced robocall annotators. Annotators perform two tasks: \textit{(a) Label Reliability (LR)} by classifying each sample as legitimate or unwanted; we compute Fleiss’ Kappa ($\kappa_f$) to measure inter-annotator agreement~\cite{fleiss1971kappa}. \textit{(b) Audio Quality (AQ)} by rating naturalness and clarity on a 10-point Mean Opinion Score (MOS) scale following ITU-T P.800~\cite{itu1996mos}; annotators also provide a 10-point confidence score. We discard 33 unwanted and 28 legitimate samples based on $\kappa_f$ and MOS with score $< 7.5$.

\textit{\textbf{Do Not Call Registry (DNCR).}} In parallel, we source unwanted robocalls drawn from the Federal Trade Commission (FTC) Do Not Call repository~\cite{robocallDataset} for real-world positive-only generalization testing purposes in an OoD fashion. The DNCR comprises real-world unwanted robocall samples in English and Mandarin.

\textit{\textbf{Data Statistics.}} We finally sourced \textbf{\texttt{Robo-SAr}} with a total of 1200 unwanted and 1202 legitimate synthesized samples across 4 different TTS with 14 different voices (male and female combined). A total of 1378 real-world unwanted robocall samples are sourced from the DNCR.

\section{Modeling}
\label{sec:modeling}

\begin{figure}
\centering
\includegraphics[width=\linewidth]{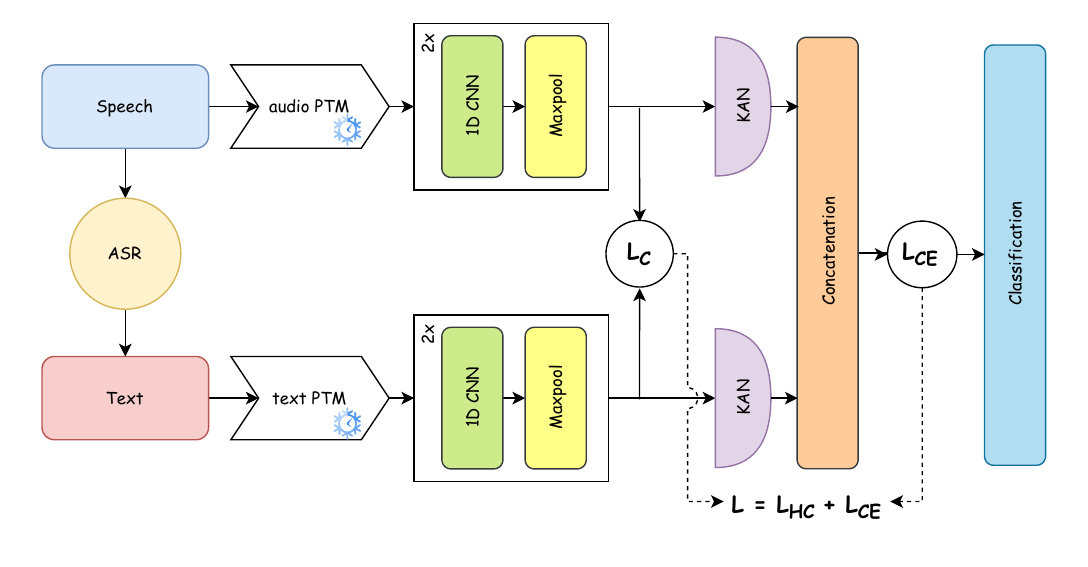}
\caption{Overview of the proposed \textbf{\texttt{RoboKA}} framework.}
\label{fig:roboka}
\end{figure}

This section describes the downstream baselines for both unimodal and multimodal settings, as well as the proposed \textbf{\texttt{RoboKA}}. An overview is provided in Fig.~\ref{fig:roboka}.

\subsection{Downstream Modeling}
\label{subsec:downstream}

Given a robocall waveform $x$, we obtain its transcript $\hat{t}=\texttt{ASR}(x)$ using OpenAI Whisper. Each modality is encoded using a frozen pre-trained model (PTM). The audio PTM produces frame-level embeddings, and the text PTM produces token-level embeddings: $h_s \in \mathbb{R}^{T_s \times d_s}$, $h_t \in \mathbb{R}^{T_t \times d_t}$, where $T_s,T_t$ denote temporal lengths and $d_s,d_t$ denote embedding dimensions.

To adapt PTM representations to robocall detection, we attach an identical lightweight convolutional head to each modality, consisting of two sequential 1D convolution layers with $64$ and $128$ filters (kernel size $3$), each followed by max pooling. To strictly enforce fixed dimensionality regardless of input duration, the final layer applies global max pooling. This yields fixed-dimensional CNN features: $u_s=f_{\text{cnn}}(h_s)$, $u_t=f_{\text{cnn}}(h_t)$, where $u_s,u_t\in\mathbb{R}^{d_u}$ and $d_u=128$. Unimodal baselines feed $u_s$ or $u_t$ into a linear or MLP classifier trained with binary cross-entropy. Multimodal baselines fuse modalities using standard operators such as concatenation, late-fusion MLPs, or cross-modal attention prior to classification.

\subsection{\textbf{\texttt{RoboKA}}}
\label{subsec:roboka}

Building on the same downstream CNN features $(u_s,u_t)$, we propose \textbf{\texttt{RoboKA}}, a KAN-informed multimodal framework that jointly learns (i) cross-modal representation alignment and (ii) discriminative robocall detection. The design is motivated by the observation that robocalls exhibit high variability in surface realization (e.g., paraphrasing, ASR noise, and prosodic manipulation), while preserving consistent cross-modal correspondences between speech and language content. \textbf{\texttt{RoboKA}} therefore enforces modality-consistent representations while learning a task-specific nonlinear decision boundary.

\paragraph{Cross-modal contrastive learning (CMCL)}

\textbf{\texttt{RoboKA}} performs CMCL at the CNN feature level. Given a minibatch of $N$ paired calls $\{(u_s^i,u_t^i)\}_{i=1}^{N}$, cosine similarity is computed as
\[
\mathrm{sim}(u_s^i,u_t^j)
=\frac{(u_s^i)^\top u_t^j}{\|u_s^i\|\,\|u_t^j\|}.
\]
We optimize an instance-level InfoNCE objective~\cite{oord2018representation}:
\[
\mathcal{L}_{C}^{x\rightarrow y}
=-\frac{1}{N}\sum_{i=1}^{N}
\log
\frac{\exp(\mathrm{sim}(u_x^i,u_y^i)/\tau)}
{\sum_{j=1}^{N}\exp(\mathrm{sim}(u_x^i,u_y^j)/\tau)},
\]
where $\tau$ is a temperature parameter, and $x$, $y$ are modalities. For inter-modal alignment, we define cross-modal contrastive loss as
$\mathcal{L}_{C}
=\frac{1}{2}\left(\mathcal{L}_{C}^{s\rightarrow t}+\mathcal{L}_{C}^{t\rightarrow s}\right)$.

Because this objective aligns paired audio and transcript representations without relying on class labels, it promotes within-call cross-modal semantic consistency while remaining robust to transcription noise and acoustic nuisance factors.

\paragraph{KAN projection and multimodal fusion}

While CNN features capture local temporal patterns, robocall detection depends on smooth nonlinear dependencies across heterogeneous acoustic and linguistic cues. To model such dependencies with a stronger inductive bias than linear or MLP heads, \textbf{\texttt{RoboKA}} adopts Kolmogorov--Arnold Networks (KANs)~\cite{liu2024kan}.

To explicitly contrast the representational capacity, consider the layer-wise operations. A standard MLP layer maps input $\mathbf{h}_{l-1}$ to $\mathbf{h}_{l}$ via a learnable linear projection $\mathbf{W}$ followed by a fixed activation $\sigma$: $\mathbf{h}_l = \sigma(\mathbf{W}_l \mathbf{h}_{l-1} + \mathbf{b}_l)$.

In contrast, a KAN layer replaces fixed non-linearities with learnable univariate functions on every edge. Given input $x\in\mathbb{R}^{d_{\text{in}}}$, a KAN layer computes
\[
\mathrm{KAN}(x)_j=\sum_{i=1}^{d_{\text{in}}}\phi_{j,i}(x_i),
\quad j=1,\dots,d_{\text{out}},
\]
where each $\phi_{j,i}(\cdot)$ is a learnable spline function parameterized as:
$\phi_{j,i}(x)
=\sum_{m=1}^{M} a^{(j,i)}_m B_m(x),$
with $B_m(\cdot)$ denoting cubic B-spline basis functions. By stacking these layers, \textbf{\texttt{RoboKA}} learns a composition of functions $\Phi = \Phi_L \circ \dots \circ \Phi_1$, enabling it to model highly non-linear cross-modal coupling that fixed-activation MLPs struggle to capture.

\paragraph{Modality-specific KAN heads}

Each modality is first processed by an independent KAN projection head:
$r_s=g^{(s)}_{\mathrm{KAN}}(u_s), \qquad
r_t=g^{(t)}_{\mathrm{KAN}}(u_t)$,
where $r_s,r_t\in\mathbb{R}^{d_k}$ and $d_k=128$. These heads act as modality-specific nonlinear calibrators, reshaping feature distributions prior to fusion without enforcing cross-modal coupling.

\paragraph{Stacked KAN-based multimodal fusion and classification}

The transformed modality features are concatenated as
$z_0=[r_s;r_t]\in\mathbb{R}^{2d_k}$.
To capture cross-modal interactions through functional composition, we apply a stacked KAN fusion head:
$z_1=g^{(f_1)}_{\mathrm{KAN}}(z_0), \qquad
\ell=g^{(f_2)}_{\mathrm{KAN}}(z_1),$
where $g^{(f_1)}_{\mathrm{KAN}}:\mathbb{R}^{2d_k}\rightarrow\mathbb{R}^{d_f}$ and
$g^{(f_2)}_{\mathrm{KAN}}:\mathbb{R}^{d_f}\rightarrow\mathbb{R}$, with $d_f=128$. The scalar $\ell$ is the output logit, and the predicted probability is $\hat{y}=\sigma(\ell)$. While a single KAN layer is additive across dimensions, stacking KAN layers introduces nonlinear coupling through composition, enabling modeling of joint audio--text dependencies.

\paragraph{Supervised detection loss}

Binary cross-entropy is used for call-level supervision:
\[
\mathcal{L}_{\mathrm{BCE}}
=-\left[y\log\hat{y}+(1-y)\log(1-\hat{y})\right],
\]
where $y\in\{0,1\}$ denotes the ground-truth label.

\paragraph{Uncertainty-based objective fusion}

Rather than manually tuning a fixed weight between contrastive alignment and classification, \textbf{\texttt{RoboKA}} adopts an uncertainty-based loss fusion strategy that adaptively balances the two objectives during training. The final objective is
\[
\mathcal{L}
=\frac{1}{2\sigma_C^2}\mathcal{L}_{C}
+\frac{1}{2\sigma_{\mathrm{BCE}}^2}\mathcal{L}_{\mathrm{BCE}}
+\log\sigma_C+\log\sigma_{\mathrm{BCE}},
\]
where $\sigma_C$ and $\sigma_{\mathrm{BCE}}$ are learnable scalar parameters capturing the relative uncertainty of the contrastive and classification objectives. This formulation enhances training stability and robustness under distribution shifts by preventing either objective from dominating the optimization.

\subsection{Training and Evaluation Protocols}

All models are trained on \textbf{\texttt{Robo-SAr}} and evaluated under four settings: T1 (TTS-engine holdout), T2 (emotion holdout), T3 (random 20\% InD holdout), and T4 (OoD positive-only testing on real-world DNCR calls). To prevent leakage, we use group-level splitting (by speaker, engine, emotion, and transcript) to ensure the training (80\%) and test (20\%) partitions are strictly disjoint. For T1-T3, we employ 5-fold cross-validation on the training set for model selection, reporting final results on the specific 20\% test holdout. For T4, models are trained on \textbf{\texttt{Robo-SAr}} using 5-fold cross-validation without access to DNCR data during training or validation, and are evaluated directly on the DNCR call set to measure OoD generalization under positive-only testing conditions.

\subsection{Experimental Results}

Table~\ref{tab:results} (a-f) reports performance for unimodal and multimodal baselines alongside the proposed \textbf{\texttt{RoboKA}} under four evaluation setups (T1--T4). A consistent modality gap emerges under distribution shift: unimodal audio PTMs perform reasonably in the easier settings but degrade under emotion-holdout and drop further on the real-world DNCR (T4), whereas unimodal text models remain comparatively stable and generalize better to DNCR. This suggests that emotional drift and channel variability primarily perturb acoustic embeddings, while transcript-derived representations are less affected by paralinguistic emotion.

Multimodal fusion consistently outperforms unimodal baselines, supporting the complementarity of audio and text. This also validates our \textbf{CMCL} objective: by explicitly encouraging audio--text agreement in the feature space, CMCL reduces modality-specific nuisance sensitivity (e.g., emotion-prosody artifacts or ASR corruption) while preserving intent-related information shared across modalities. The audio stream retains paralinguistic cues (e.g., urgency, stress) that remain informative when transcripts are imperfect, while text anchors semantic intent when acoustic patterns shift.

This complementarity is reflected in the fusion hierarchy. While standard operators such as concatenation ($+$), late-fusion MLPs ($\oplus$), and cross-modal attention ($\otimes$) provide gains, \textbf{\texttt{RoboKA}} ($\boxplus$) yields the most consistent improvements, particularly in the hardest real-world setting (DNCR). On T4, the strongest non-\textbf{\texttt{RoboKA}} baseline reaches \textbf{67.21} uRc, whereas \textbf{\texttt{RoboKA}} achieves \textbf{82.21} uRc (\textbf{+15.00}). Similarly, \textbf{\texttt{RoboKA}} improves T1 macro-F1 from \textbf{62.97} to \textbf{80.41} (\textbf{+17.44}) and T2 macro-F1 from \textbf{67.97} to \textbf{76.73} (\textbf{+8.76}). These gains support our hypothesis that robust robocall detection requires conditional feature-wise calibration and stable evidence aggregation. Unlike MLP fusion, which can co-adapt brittle cross-modal shortcuts, KAN-based projection and fusion learn controlled nonlinear gating that suppresses unreliable dimensions and sharpens the unwanted/legitimate boundary under shifts. The robustness across diverse shifts further supports the effectiveness of the uncertainty-weighted objective fusion, which adaptively balances $\mathcal{L}_{C}$ and $\mathcal{L}_{BCE}$ to prevent either objective from dominating optimization when one modality becomes unreliable.

Overall, these results indicate that robust unwanted-call detection benefits from both \textit{what is said} and \textit{how it is said}, and that principled alignment (CMCL), calibration-centric fusion (KAN), and adaptive optimization (Uncertainty) are critical for resilience under adversarial and real-world shifts.

\begin{table}
\centering
\caption{Empirical results leveraging both audio and text modalities in unimodal and multimodal settings. Here, mRC, mF1, and uRc define macro Recall, macro F1, and recall for unwanted samples, respectively. For Ablation, we show the best results for each modality and objective functions.}
\label{tab:results}
\scriptsize
\renewcommand{\arraystretch}{1.15}
\setlength{\tabcolsep}{3.2pt}
\begin{tabular}{l|cc|cc|cc|c}
\toprule
\textbf{Model} 
& \multicolumn{2}{c|}{\textbf{T1}} 
& \multicolumn{2}{c|}{\textbf{T2}} 
& \multicolumn{2}{c|}{\textbf{T3}}
& \textbf{T4} \\
\cline{2-8}
& mRc & mF1 & mRc & mF1 & mRc & mF1 & uRc \\
\midrule

\multicolumn{8}{c}{(a) \textbf{Unimodal (Audio)}} \\
\midrule
Wav2Vec2
& 57.02 & 56.61 & 51.48 & 51.07 & 95.19 & 95.27 & 45.83 \\
WavLM
& \textbf{60.03} & \textbf{59.58} & \textbf{54.52} & \textbf{54.19} & 98.15 & 98.18 & \textbf{46.62} \\
HuBERT
& 59.03 & 58.74 & 54.47 & 54.08 & \textbf{99.31} & \textbf{99.36} & 45.88 \\
\midrule

\multicolumn{8}{c}{(b) \textbf{Unimodal (Text)}} \\
\midrule
RoBERTa
& 56.41 & 56.02 & 60.94 & 60.49 & 95.02 & 95.18 & 56.84 \\
GPT2
& 55.06 & 55.61 & 61.87 & 62.79 & 96.18 & 97.12 & 56.52 \\
BERT
& \textbf{60.02} & \textbf{60.16} & \textbf{65.03} & \textbf{65.09} & \textbf{99.25} & \textbf{99.29} & \textbf{58.13} \\
\midrule

\multicolumn{8}{c}{(c) \textbf{Multimodal $\rightarrow$ Concat ($+$)}} \\
\midrule
Wav2Vec2 $+$ RoBERTa
& 56.44 & 56.67 & 61.88 & 61.97 & 96.11 & 96.24 & 57.86 \\
Wav2Vec2 $+$ GPT2
& 57.11 & 57.39 & 62.27 & 62.58 & 96.40 & 96.55 & 57.55 \\
Wav2Vec2 $+$ BERT
& 62.01 & 62.22 & 65.33 & 65.54 & 97.61 & 97.70 & 58.71 \\
WavLM$+$ RoBERTa
& 60.09 & 60.33 & 65.01 & 65.17 & 98.71 & 98.82 & 59.47 \\
WavLM$+$ GPT2
& 59.38 & 59.71 & 64.55 & 64.69 & 98.80 & 98.90 & 59.62 \\
WavLM$+$ BERT
& 61.47 & 61.86 & 66.07 & 66.34 & 99.19 & 99.26 & 64.69 \\
HuBERT$+$ RoBERTa
& 60.99 & 61.11 & 65.73 & 65.89 & 99.18 & 99.23 & 59.04 \\
HuBERT$+$ GPT2
& 60.27 & 60.41 & 65.38 & 65.63 & 99.20 & 99.26 & 59.11 \\
HuBERT$+$ BERT
& \textbf{62.55} & \textbf{62.68} & \textbf{67.11} & \textbf{67.28} & \textbf{99.44} & \textbf{99.48} & \textbf{65.27} \\
\midrule

\multicolumn{8}{c}{(d) \textbf{Multimodal $\rightarrow$ Late-Fusion MLP ($\oplus$)}} \\
\midrule
Wav2Vec2 $\oplus$ RoBERTa
& 57.01 & 57.28 & 62.57 & 62.94 & 96.52 & 96.66 & 58.44 \\
Wav2Vec2 $\oplus$ GPT2
& 57.39 & 57.66 & 62.79 & 63.06 & 96.75 & 96.89 & 58.21 \\
Wav2Vec2 $\oplus$ BERT
& 60.84 & 60.99 & 65.87 & 65.98 & 98.01 & 98.08 & 59.53 \\
WavLM$\oplus$ RoBERTa
& 60.96 & 61.09 & 66.08 & 66.21 & 98.92 & 99.02 & 60.11 \\
WavLM$\oplus$ GPT2
& 60.14 & 60.29 & 65.93 & 66.07 & 98.98 & 99.08 & 60.03 \\
WavLM$\oplus$ BERT
& 61.61 & 61.74 & 66.71 & 66.88 & 99.32 & 99.38 & 66.02 \\
HuBERT$\oplus$ RoBERTa
& 61.77 & 61.86 & 66.95 & 67.07 & 99.34 & 99.38 & 59.94 \\
HuBERT$\oplus$ GPT2
& 61.26 & 61.39 & 66.54 & 66.69 & 99.36 & 99.41 & 59.86 \\
HuBERT$\oplus$ BERT
& \textbf{62.64} & \textbf{62.72} & \textbf{67.32} & \textbf{67.46} & \textbf{99.56} & \textbf{99.60} & \textbf{66.58} \\
\midrule

\multicolumn{8}{c}{(e) \textbf{Multimodal $\rightarrow$ Cross-Modal Attention ($\otimes$)}} \\
\midrule
Wav2Vec2 $\otimes$ RoBERTa
& 57.48 & 57.73 & 62.87 & 63.16 & 97.05 & 97.18 & 59.14 \\
Wav2Vec2 $\otimes$ GPT2
& 57.83 & 58.07 & 63.01 & 63.33 & 97.20 & 97.33 & 58.88 \\
Wav2Vec2 $\otimes$ BERT
& 61.32 & 61.49 & 66.28 & 66.44 & 98.35 & 98.41 & 60.24 \\
WavLM$\otimes$ RoBERTa
& 61.33 & 61.48 & 66.61 & 66.81 & 99.18 & 99.24 & 61.06 \\
WavLM$\otimes$ GPT2
& 60.73 & 60.88 & 66.17 & 66.32 & 99.22 & 99.28 & 60.95 \\
WavLM$\otimes$ BERT
& \celllb{62.81} & \celllb{62.97} & 67.41 & 67.59 & 99.52 & 99.56 & 66.93 \\
HuBERT$\otimes$ RoBERTa
& 62.04 & 62.18 & 67.11 & 67.26 & 99.46 & 99.50 & 61.55 \\
HuBERT$\otimes$ GPT2
& 61.67 & 61.81 & 66.92 & 67.09 & 99.48 & 99.52 & 61.33 \\
HuBERT$\otimes$ BERT
& 62.71 & 62.82 & \celllb{67.88} & \celllb{67.97} & \celllb{99.66} & \celllb{99.69} & \celllb{67.21} \\
\midrule

\multicolumn{8}{c}{(f) \textbf{Multimodal $\rightarrow$ \textbf{\texttt{RoboKA}} ($\boxplus$)}} \\
\midrule
Wav2Vec2 $\boxplus$ RoBERTa
& 76.33 & 76.57 & 73.61 & 73.89 & 97.70 & 97.82 & 74.08 \\
Wav2Vec2 $\boxplus$ GPT2
& 76.52 & 76.75 & 73.41 & 73.66 & 97.85 & 97.96 & 73.55 \\
Wav2Vec2 $\boxplus$ BERT
& 78.09 & 78.27 & 74.56 & 74.73 & 98.70 & 98.74 & 75.72 \\
WavLM$\boxplus$ RoBERTa
& 78.51 & 78.66 & 75.02 & 75.19 & 99.45 & 99.49 & 79.44 \\
WavLM$\boxplus$ GPT2
& 77.93 & 78.08 & 74.41 & 74.58 & 99.48 & 99.52 & 78.86 \\
WavLM$\boxplus$ BERT
& 79.86 & 79.98 & 76.22 & 76.34 & 99.70 & 99.73 & 81.03 \\
HuBERT$\boxplus$ RoBERTa
& 80.05 & 80.17 & \celldb{76.58} & \celldb{76.73} & 99.60 & 99.63 & \celldb{82.21} \\
HuBERT$\boxplus$ GPT2
& 79.59 & 79.71 & 76.01 & 76.16 & 99.62 & 99.66 & 81.49 \\
HuBERT$\boxplus$ BERT
& \celldb{80.34} & \celldb{80.41} & 76.52 & 76.63 & \celldb{99.78} & \celldb{99.80} & 81.74 \\
\midrule

\multicolumn{8}{c}{(g) \textbf{Ablations (Task-specific Best-Results)}} \\
\midrule
A (KAN, $\mathcal{L}_{BCE}$) 
& 66.11 & 65.24 & 61.08 & 60.41 & 97.12 & 97.26 & 51.88 \\
T (KAN, $\mathcal{L}_{BCE}$)
& 68.42 & 67.77 & 66.95 & 67.21 & 98.04 & 98.19 & 63.73 \\
A$+$T (KAN, $\mathcal{L}_{BCE}$) 
& 71.36 & 71.74 & 69.02 & 69.38 & 98.61 & 98.74 & 68.95 \\
A$\oplus$T (no KAN, $\mathcal{L}_C+\mathcal{L}_{BCE}$) 
& 72.09 & 72.44 & 70.08 & 70.37 & 98.83 & 98.96 & 70.51 \\
A$\oplus$T (KAN, $\mathcal{L}_C+\mathcal{L}_{BCE}$) 
& \celllb{74.88} & \celllb{75.19} & \celllb{72.41} & \celllb{72.76} & \celllb{99.11} & \celllb{99.24} & \celllb{73.62} \\
A$\oplus$T (no KAN, $\mathcal{L}$) 
& 72.61 & 72.94 & 71.36 & 70.69 & 99.05 & 99.18 & 70.88 \\
A$\boxplus$T (KAN, $\mathcal{L}$) (\textbf{\texttt{RoboKA}})
& \celldb{80.34} & \celldb{80.41} & \celldb{76.58} & \celldb{76.73} & \celldb{99.78} & \celldb{99.80} & \celldb{81.74} \\
\bottomrule

\end{tabular}
\end{table}

\subsection{Ablation of \textbf{\texttt{RoboKA}}}

Table~\ref{tab:results} (g) presents a joint task-wise and component-wise ablation of \textbf{\texttt{RoboKA}}, showing that audio and text provide complementary but unevenly reliable cues under distribution shift, which motivates multimodal learning. Moving from unimodal to multimodal training yields clear gains, and adding cross-modal contrastive learning further improves robustness by encouraging modality agreement beyond supervised classification. Removing KAN while maintaining the same objectives consistently degrades performance, with the largest drops occurring in the most challenging settings (notably DNCR), indicating that KAN-based feature-wise nonlinear calibration is more robust than standard fusion heads. Finally, uncertainty-weighted training improves stability and generalization across shifts, and the full \textit{A$\boxplus$T (KAN, $\mathcal{L}$)} configuration achieves the best overall performance, validating the combined role of multimodal evidence, alignment, calibration-centric fusion, and adaptive objective balancing.



\section{Discussion}
\label{sec:discussion}

Our results highlight a robustness gap under distribution shift: audio PTMs are fragile to paralinguistic and channel variations (e.g., emotion, recording quality), whereas text representations remain stable but are vulnerable to ASR errors. Multimodal fusion bridges this by leveraging complementary evidence. \textbf{\texttt{RoboKA}} yields the most consistent improvements by coupling (i) CMCL to enforce intent-based feature alignment, (ii) Stacked KAN fusion layers to capture complex non-linear interactions via learnable splines, and (iii) Uncertainty-weighted objective fusion to adaptively balance alignment and classification losses.

\noindent \textit{Limitations:} First, our study is restricted primarily to the English language only, leaving multilingual generalization unexplored. Second, while our real-world evaluation on DNRC contains only unwanted robocalls and lacks publicly available legitimate datasets. This leads to a recall-only analysis, as false positives or precision in this setting remains invalid. Third, despite modeling adversarial factors such as voice cloning and emotion conditioning, a residual domain gap between synthetic telephony simulations and real-world acoustic conditions may affect generalization and motivate future adaptation or fine-tuning.

\section{Conclusion}
\label{sec:conclusion}

We introduced \textbf{\texttt{Robo-SAr}}, an adversarial robocall benchmark for robocall surveillance research, spanning transcript manipulation, emotion elicitation, and voice cloning, and proposed \textbf{\texttt{RoboKA}}, a KAN-informed multimodal framework for robust unwanted-call detection. Experiments across four tasks, TTS-holdout, emotion-holdout, 20\%-holdout, and real-world DNCR, show that unimodal detectors are brittle under realistic shifts, whereas principled multimodal learning improves generalization. By integrating CMCL for feature alignment, stacked KAN fusion for non-linear coupling, and uncertainty-aware objective balancing, \textbf{\texttt{RoboKA}} consistently outperforms strong unimodal and multimodal baselines, supporting the need to model both \emph{what is said} and \emph{how it is said} for robust robocall defense.

\bibliographystyle{IEEEbib}
\bibliography{icme2025references}

\end{document}